\documentclass{article}
\usepackage{spconf,amsmath,graphicx}

\usepackage{enumitem}
\setlist{nosep, leftmargin=14pt}

\usepackage{mwe} % to get dummy images
\usepackage{multirow}
\usepackage{amssymb}
\usepackage{algorithmic}
\usepackage{algorithm}
\usepackage{subfig}
\usepackage{graphicx}
\usepackage[colorlinks=true, urlcolor=blue, linkcolor=red]{hyperref}
\usepackage{color, colortbl}

\definecolor{Worse}{rgb}{0.9,0.9,1}
\definecolor{Better}{rgb}{1,0.9,0.9}
\title{IPMN Risk Assessment under Federated Learning Paradigm}
\name{\begin{tabular}{c}Hongyi Pan$^{1*}$, Ziliang Hong$^{1*}$, Gorkem Durak$^1$, Elif Keles$^1$,  Halil Ertugrul Aktas$^1$, Yavuz Taktak$^2$,\\Alpay Medetalibeyoglu$^3$, Zheyuan Zhang$^1$,
Yury Velichko$^1$, Concetto Spampinato$^4$, Ivo Schoots$^5$, \\Marco J. Bruno$^6$, Pallavi Tiwari$^7$, Candice Bolan$^8$, Tamas Gonda$^9$, Frank Miller$^1$, \\Rajesh N. Keswani$^{10}$, Michael B. Wallace$^8$, Ziyue Xu$^{11}$, Ulas Bagci$^1$\end{tabular}\thanks{$^{*}$These authors contributed equally.}
}
\address{$^1$Department of Radiology, Northwestern University, Chicago, IL, USA\\
$^2$Department of Radiology, Istanbul Faculty of Medicine, Istanbul University, Istanbul, Turkey\\
$^3$Department of Internal Medicine, Istanbul Faculty of Medicine, Istanbul University, Istanbul, Turkey\\
$^4$Department of Electrical, Electronic and Computer Engineering, University of Catania, Catania, Italy \\
$^5$Department of Radiology and Nuclear Medicine, Erasmus Medical Center, Rotterdam, Netherlands\\
$^6$Department of Gastroenterology and Hepatology, Erasmus Medical Center, Rotterdam, Netherlands\\
$^7$Department of Biomedical Engineering and Radiology, University of Wisconsin, Madison, WI, USA\\
$^8$Division of Gastroenterology and Hepatology, Mayo Clinic, Jacksonville, FL, USA\\
$^9$Division of Gastroenterology and Hepatology, New York University, New York, NY, USA\\
$^{10}$Department of Gastroenterology and Hepatology, Northwestern University, Chicago, IL, USA\\
$^{11}$NVIDIA, Bethesda, MD, USA
}
\begin{document}
\ninept
\maketitle
\begin{abstract}
Accurate classification of Intraductal Papillary Mucinous Neoplasms (IPMN) is essential for identifying high-risk cases that require timely intervention. In this study, we develop a federated learning framework for multi-center IPMN classification utilizing a comprehensive pancreas MRI dataset. This dataset includes 652 T1-weighted and 655 T2-weighted MRI images, accompanied by corresponding IPMN risk scores from 7 leading medical institutions, making it the largest and most diverse dataset for IPMN classification to date. We assess the performance of DenseNet-121 in both centralized and federated settings for training on distributed data. Our results demonstrate that the federated learning approach achieves high classification accuracy comparable to centralized learning while ensuring data privacy across institutions. This work marks a significant advancement in collaborative IPMN classification, facilitating secure and high-accuracy model training across multiple centers.

\end{abstract}
\begin{keywords}
Pancreas MRI, IPMN classification, federated learning, multi-center
\end{keywords}

\section{Introduction}
Intraductal papillary mucinous neoplasms (IPMNs) are cystic lesions in the pancreas that can evolve into malignant tumors, necessitating timely and accurate classification for effective patient management. Magnetic resonance imaging (MRI) is a critical tool for visualizing IPMNs, enabling radiologists to assess features that inform clinical decisions~\cite{hussein2018deep,lalonde2019inn,salanitri2022neural}. However, the variability in imaging protocols across institutions and the rarity of IPMNs make it challenging to develop robust, generalizable machine learning models for IPMN classification. Accessing diverse, large-scale datasets is essential for improving model accuracy, yet centralized data aggregation is often unfeasible due to privacy regulations surrounding medical imaging.

Federated learning (FL) has emerged as a promising solution for collaborative model training across institutions while preserving patient data privacy~\cite{mcmahan2017communication,li2020federated, wang2020tackling, li2020review,zhang2021survey,guo2022auto,collins2022fedavg,yuan2022convergence,jiang2023fair,gholami2024digest,miao2024federated,pan2024frequency}. By sharing model updates instead of raw data, federated learning allows for multi-center collaborations without exposing patient information. This study builds upon this framework by introducing the largest-ever multi-center IPMN classification dataset, encompassing MRI scans from multiple institutions. This extensive dataset enables a comprehensive evaluation of IPMN classification performance, comparing traditional centralized learning approaches with federated learning in a multi-center setting.

Our contribution can be summarized as the following: (1) we present a multi-center dataset for IPMN classification, providing a benchmark for future research; (2) we perform an extensive evaluation of both centralized and federated learning approaches across multiple institutions; and (3) we demonstrate the efficacy of federated learning in achieving high classification accuracy while maintaining patient privacy. By addressing these key challenges, this study offers a scalable and privacy-preserving approach to IPMN classification in pancreas MRI, paving the way for more effective clinical applications in the future.

\section{Methodology}
\subsection{Data Collection and Preprocessing}

\begin{table*}[tb]
\setlength{\tabcolsep}{3pt}
\caption{Pancreases Data Distribution.}\vspace{-20pt}
\begin{center}
\begin{tabular}{ll|cccc|cccc}
\hline
&\multirow{2}{*}{\textbf{Data Centers}}&\multicolumn{4}{c|}{\textbf{T1-Weighted Modality}}&\multicolumn{4}{c}{\textbf{T2-Weighted Modality}}\\
&&No Risk&Low Risk& High Risk&Total&No Risk&Low Risk& High Risk&Total\\
\hline
1&New York University Langone Health (NYU) & 48 & 79 & 23 & 150 & 48 & 79 & 24 & 151\\
2&Mayo Clinic Florida (MCF) & 29 & 42 & 63 & 134 & 25 & 42 & 63 & 130\\
3&Northwestern University (NU) & 43 & 126 & 17 & 186 & 44 & 127 & 16 & 187\\
4&Allegheny Health Network (AHN) & 1 & 11 & 4 & 16 & 1 & 13 & 4 & 18\\
5&Mayo Clinic Arizona (MCA) & 0 & 10 & 14 & 24 & 0 & 7 & 16 & 23\\
6&Istanbul University Faculty of Medicine (IU) &  3 & 48 & 13 & 64 & 3 & 46 & 14 & 63\\
7&Erasmus Medical Center (EMC) &  40 & 23 & 15 & 78 & 38 & 30 & 15 & 83\\
\hline
&\textbf{Total} & 164 & 339 & 149 & 652 & 159 & 344 & 152 & 655\\
\hline
\end{tabular}\vspace{-10pt}
\label{tab: Distribution}
\end{center}
\end{table*}

\begin{figure*}[htbp]
\centering
\subfloat[NYU]{
\begin{minipage}{0.1\linewidth}
\includegraphics[width=\linewidth, height=\linewidth]{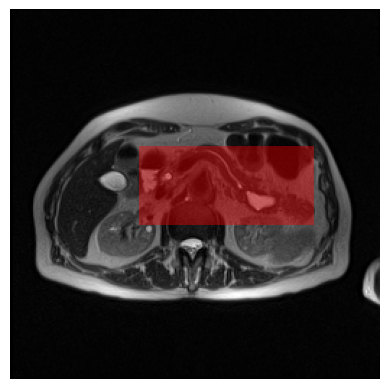}\\
\includegraphics[width=\linewidth, height=\linewidth]{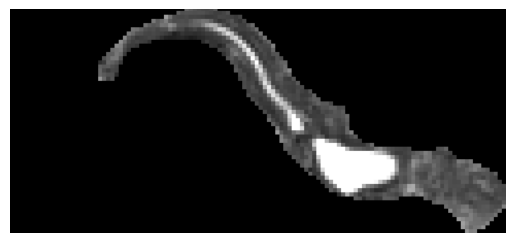}
\end{minipage}}
\subfloat[MCF]{
\begin{minipage}{0.1\linewidth}
\includegraphics[width=\linewidth, height=\linewidth]{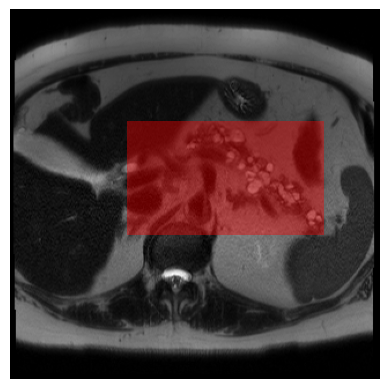}\\
\includegraphics[width=\linewidth, height=\linewidth]{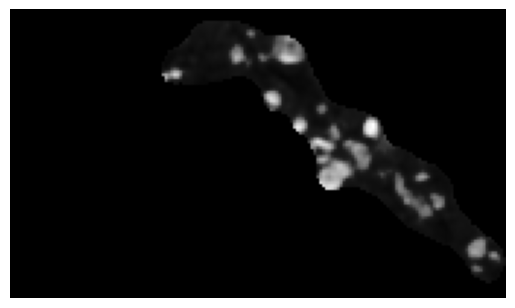}
\end{minipage}}
\subfloat[NU]{
\begin{minipage}{0.1\linewidth}
\includegraphics[width=\linewidth, height=\linewidth]{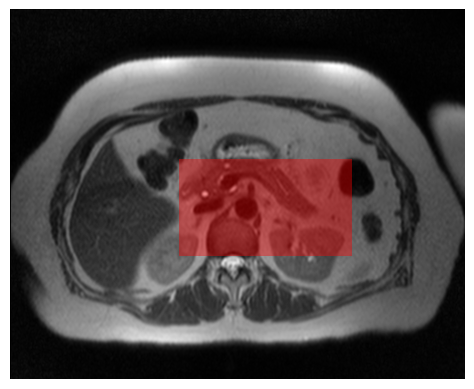}\\
\includegraphics[width=\linewidth, height=\linewidth]{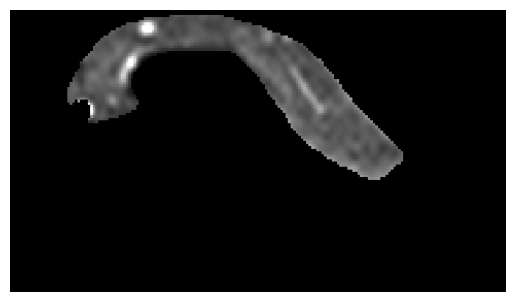}
\end{minipage}}
\subfloat[AHN]{
\begin{minipage}{0.1\linewidth}
\includegraphics[width=\linewidth, height=\linewidth]{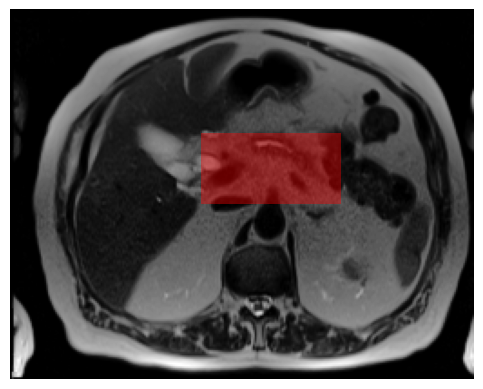}\\
\includegraphics[width=\linewidth, height=\linewidth]{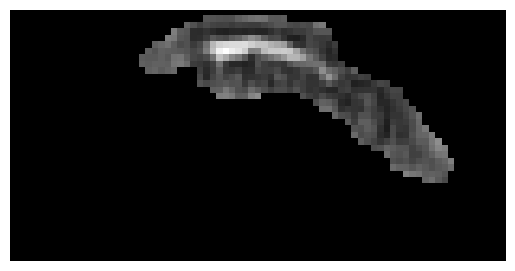}
\end{minipage}}
\subfloat[MCA]{
\begin{minipage}{0.1\linewidth}
\includegraphics[width=\linewidth, height=\linewidth]{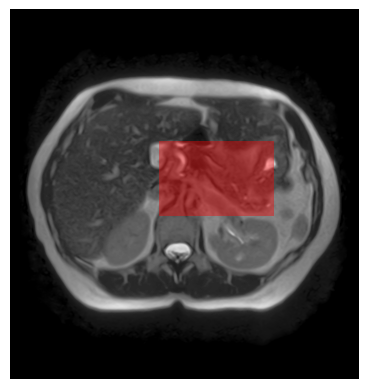}\\
\includegraphics[width=\linewidth, height=\linewidth]{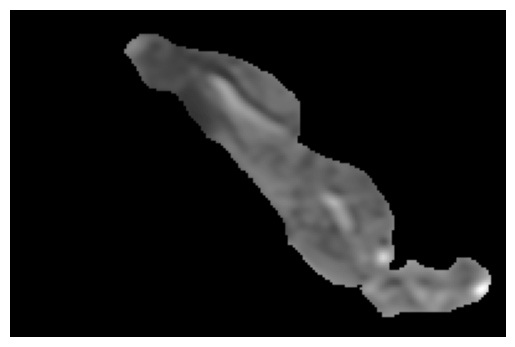}
\end{minipage}}
\subfloat[IU]{
\begin{minipage}{0.1\linewidth}
\includegraphics[width=\linewidth, height=\linewidth]{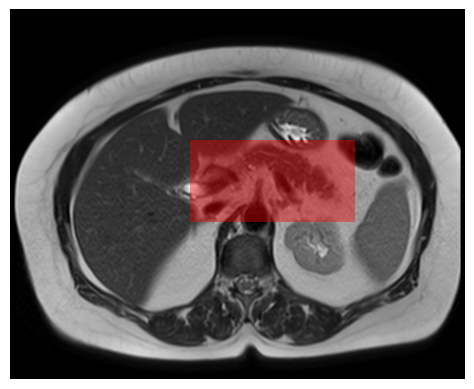}\\
\includegraphics[width=\linewidth, height=\linewidth]{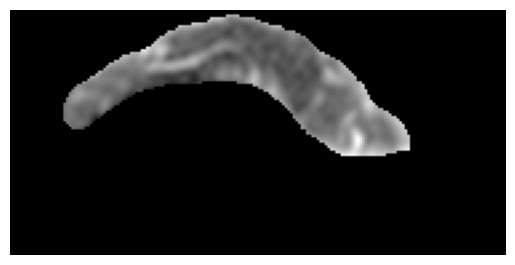}
\end{minipage}}
\subfloat[EMC]{
\begin{minipage}{0.1\linewidth}
\includegraphics[width=\linewidth, height=\linewidth]{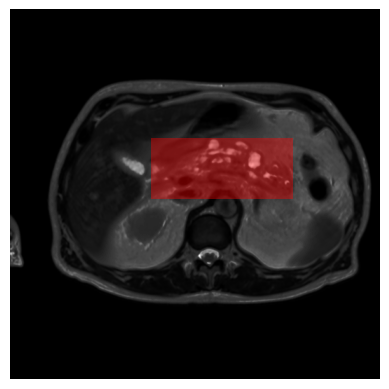}\\
\includegraphics[width=\linewidth, height=\linewidth]{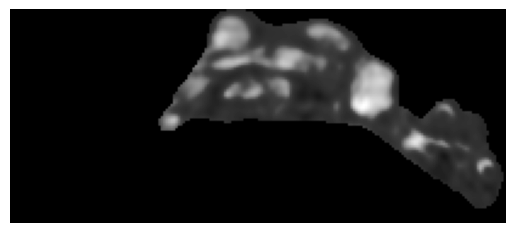}
\end{minipage}}\vspace{-10pt}
\caption{MRI images from different centers, highlighting the corresponding regions of interest (ROIs). The top panel displays the MRI images with the ROIs indicated, while the bottom panel shows the cropped ROIs for closer examination.}\vspace{-10pt}
\label{fig: samples}
\end{figure*}

This study utilizes an extended version of the pancreas MRI dataset introduced in \cite{zhang2025large} to identify high-risk Intraductal Papillary Mucinous Neoplasms (IPMN) cases. The dataset comprises 723 T1-weighted and 738 T2-weighted MRI images collected from 7 medical centers: New York University Langone Health (NYU), Mayo Clinic Florida (MCF), Northwestern University (NU), Allegheny Health Network (AHN), Mayo Clinic Arizona (MCA), Istanbul University Faculty of Medicine (IU), and Erasmus Medical Center (EMC). For IPMN classification, we exclude images without necessary pathological information. Consequently, a total of 652 T1-weighted and 655 T2-weighted images relevant to IPMN risk analysis are included. These images are cropped to focus on regions of interest (ROI) and categorized into no-risk, low-risk, or high-risk groups. The distribution of images across centers is presented in Table~\ref{tab: Distribution}, and sample MRI images are displayed in Fig.~\ref{fig: samples}. The dataset is available on OSF\footnote{\url{https://osf.io/74vfs/}}.

\subsection{3D DenseNet for Pancreas MRI IPMN Classification}
For pancreas MRI IPMN classification, we employ DenseNet-121 (Densely Connected Convolutional Network)~\cite{huang2017densely} as the backbone of our 3D convolutional model. DenseNet is a deep learning architecture that connects each layer to all preceding layers within dense blocks, allowing each layer to receive cumulative feature information from previous layers. Unlike traditional networks, where information flows sequentially, DenseNet’s dense connections enable extensive feature reuse and strengthen gradient flow, enhancing model performance with fewer parameters. Dense blocks are separated by transition layers, which downsample feature maps to maintain model compactness and efficiency. DenseNet’s design mitigates the vanishing gradient problem and improves parameter efficiency, making it particularly effective for complex image classification tasks.

\subsection{Federated Multi-Center IPMN  Classification}
To enable privacy-preserving, collaborative learning across multiple institutions, we adopt a federated learning framework for training the IPMN classification model on pancreas MRI data. By keeping patient data localized at each institution, federated learning ensures that sensitive medical information remains private while allowing for the joint optimization of a global model. We employ two federated optimization algorithms—Federated Averaging (FedAvg)~\cite{mcmahan2017communication} and Federated Proximal (FedProx)~\cite{li2020federated}—to facilitate training and address challenges associated with data heterogeneity across sites.

\subsubsection{FedAvg}
Federated Averaging (FedAvg)~\cite{mcmahan2017communication} is the baseline aggregation method used in this study. In the FedAvg method, each institution $k$ independently trains a local model $\mathbf{w}_k$ on its dataset for a fixed number of epochs. Afterward, each local model's parameters are transmitted to a central server, which computes the global model $\mathbf{w}$ by averaging the updates from all institutions.

In detail, the FedAvg with $K$ institutions can be implemented in the following two parts:
\begin{enumerate}
    \item \textbf{Local Update}: Each institution $k$ minimizes its local objective $F_k(\mathbf{w})$, defined as:
    \begin{equation}
        F_k(\mathbf{w}) = \frac{1}{N_k} \sum_{i \in \mathcal{D}_k} \ell(f(\mathbf{x}_i; \mathbf{w}), \mathbf{y}_i),
    \end{equation}
    where $N_k$ is the number of samples at institution $k$, $\mathcal{D}_k$ is the dataset for institution $k$, $\ell(\cdot)$ is the loss function (\textit{e.g.}, cross-entropy for classification), and $f(\mathbf{x}_i; \mathbf{w})$ computes model output for input $\mathbf{x}_i$ with parameters $\mathbf{w}$.
    
    \item \textbf{Global Aggregation}: The central server aggregates local models by weighted averaging:
    \begin{equation}
        \mathbf{w} = \frac{\sum_{k=0}^{K-1} N_k \mathbf{w}_k}{\sum_{k=0}^{K-1} N_k},
    \end{equation}
    where $\mathbf{w}_k$ is the locally updated model from institution $k$.
\end{enumerate}

FedAvg is computationally efficient and performs well under homogeneous data conditions. However, data variations across institutions can sometimes degrade model performance.

\subsubsection{FedProx}
To handle data heterogeneity across institutions, we incorporate the Federated Proximal (FedProx) algorithm~\cite{li2020federated}, which modifies the local objective by adding a proximal term that regularizes the local updates, keeping them close to the global model $\mathbf{w}$:
    \begin{equation}
        F_k^{\text{prox}}(\mathbf{w}) = F_k(\mathbf{w}) + \frac{\mu}{2} \| \mathbf{w} - \mathbf{w}_t \|^2,
    \end{equation}
    where $\mathbf{w}_t$ is the global model from the previous communication round, $\mu \geq 0$ is a tuning parameter that controls the regularization strength, and $\| \cdot \|$ denotes the $L^2$-norm. The proximal term $\frac{\mu}{2} \| \mathbf{w} - \mathbf{w}_t \|^2$ penalizes large deviations from the global model, improving stability in non-iid settings. This modification helps mitigate variability due to non-iid data distributions among institutions.

FedProx is designed to address FedAvg's limitations when dealing with data and system heterogeneity across clients. However, FedProx can perform worse than FedAvg because its regularization term may overly restrict local updates, slowing convergence and reducing accuracy, especially if client data is only mildly heterogeneous. Its performance also heavily depends on tuning a sensitive hyperparameter and requires more computation, which can hinder efficiency in resource-constrained settings. Therefore, we implement two federated learning methods in this study to present a benchmark on our proposed multi-center pancreas MRI IPNM dataset.

\subsubsection{Training Protocol}

Institutions independently train a 3D DenseNet-121 on their local MRI data in each communication round, optimizing the FedAvg or FedProx objective. The central server iteratively aggregates the local models, and the process continues until convergence or for a predetermined number of rounds. We evaluate the classification performance and generalizability of the model trained with each algorithm, comparing these results with a centralized baseline.

\section{Experimental Results}
We begin this section by implementing a 3-class classification experiment to distinguish between no-risk, low-risk, and high-risk IPMN cases using various 3D deep neural network architectures. This initial experiment is conducted without applying the multi-center federated learning setting.

Subsequently, we conduct a binary classification experiment (high-risk vs. non-high-risk) using the best-performing model (DenseNet-121) from the initial 3-class experiment, evaluating its effectiveness in a multi-center setting with both centralized and federated learning approaches. The 2-class setup is chosen because our primary objective is to identify high-risk IPMN cases. Additionally, some centers have few or no no-risk cases, making 3-class classification less feasible in the federated setting.

All images cropped to ROI are resized to $96 \times 96 \times 96$. For patient-level classification, we employ the Medical Open Network for Artificial Intelligence (MONAI) framework~\cite{cardoso2022monai} to build 3D versions of the models, training them separately on T1-weighted and T2-weighted datasets. The models are optimized by minimizing cross-entropy loss, and we retain the model checkpoint with the highest global area under the curve (AUC). Model performance is evaluated based on accuracy (ACC) and AUC. To ensure robust results, cross-validation is applied across experiments, with each performance metric reported as the mean $\pm$ standard deviation. All experiments are conducted using PyTorch~\cite{paszke2019pytorch}.

\subsection{IPMN MRI 3-Class Classification}
In this study, we evaluate 6 state-of-the-art convolutional neural networks (CNNs): ResNet-34~\cite{he2016deep}, V2~\cite{sandler2018mobilenetv2}, ShuffleNet-V2~\cite{ma2018shufflenet}, EfficientNet-B0~\cite{tan2019efficientnet}, MobileNet-and DenseNet-121~\cite{huang2017densely}. The training process is conducted over 100 epochs using the AdamW optimizer~\cite{loshchilov2017decoupled}, with an initial learning rate of 0.001, a batch size of 16, and a weight decay of 0.01. The learning rate is reduced by a factor of 10 every 30 epochs. 5-fold cross-validation is implemented to ensure the robustness of the results. 

Table~\ref{tab: 3class} presents the 3-class classification results for IPMN MRI, with MACs and parameters calculated using the ``ptflops'' library\footnote{\url{https://github.com/sovrasov/flops-counter.pytorch}}. %Notably, ResNet-34 has more MACs and parameters than ResNet-50, as it has more $3 \times 3 \times 3$ Conv3D layers, which are computationally intensive in the 3D model architecture. 
DenseNet-121 exhibits the highest AUC results, achieving 0.7632 for T1-weighted modality and 0.8092 for T2-weighted modality. This model strikes a balance with a moderate parameter scale while demonstrating relatively strong performance. In contrast, ResNet-34 achieves slightly lower AUC values of 0.7372 on T1-weighted modality and 0.7909 on T2-weighted modality. %however, they require significantly larger FLOPS and have more parameters compared to DenseNet-121.

The performance of the more lightweight models shows a declining trend as the number of parameters decreases. For example, ShuffleNet-V2 achieves AUC values of 0.6929 and 0.7617. This trend indicates that the performance of these lighter-weight models tends to decline with fewer parameters.

\begin{table*}[htbp]
\caption{IPMN MRI 3-Class Classification Results.}
\begin{center}
\begin{tabular}{l|cc|cc|cc}
\hline
\multirow{2}{*}{\textbf{Model}} & \textbf{MACs} & \textbf{Parameters} & \multicolumn{2}{c|}{\textbf{T1-Weighted Modality}} & \multicolumn{2}{c}{\textbf{T2-Weighted Modality}}  \\
 & \textbf{(G)}& \textbf{(M)} &\textbf{ACC} & \textbf{AUC} & \textbf{ACC} & \textbf{AUC} \\
\hline
ResNet-34 & 183.04 &  63.47 &56.62$\pm$9.13&72.01$\pm$5.27&58.57$\pm$3.81&78.63$\pm$2.04\\
% ResNet-50 & 138.07  & 46.17& 63.62$\pm$3.89 &78.01$\pm$3.66& 61.48$\pm$6.13 & 81.46$\pm$1.97\\
ShuffleNet-V2 & 0.57 & 1.3  &46.65$\pm$9.03 & 65.63$\pm$3.80&58.49$\pm$3.82 & 71.50$\pm$3.08\\
MobileNet-V2 & 2.06 & 2.36 &49.52$\pm$5.30&65.49$\pm$4.64&45.48$\pm$7.88&65.76$\pm$2.25\\
EfficientNet-B0 & 1.21 & 4.69&55.49$\pm$2.75&68.07$\pm$1.57&59.37$\pm$5.78&74.41$\pm$4.10\\
DenseNet-121 & 18.31 & 11.25& \bf{58.69$\pm$4.28} & \bf{75.59$\pm$7.05}& \bf{65.89$\pm$4.98} & \bf{81.09$\pm$4.10}\\
\hline
\end{tabular}
\label{tab: 3class}
\end{center}
\end{table*}

\begin{table*}[htbp]
\caption{Binary Classification Results for Multi-Center IPMN MRI Using DenseNet-121.}
\begin{center}
\begin{tabular}{l|cc|cc}
\hline
\multirow{2}{*}{\textbf{Method}}&\multicolumn{2}{c|}{\textbf{T1-Weighted Modality}}&\multicolumn{2}{c}{\textbf{T2-Weighted Modality}}\\
&\textbf{ACC}  & \textbf{AUC}& \textbf{ACC} & \textbf{AUC} \\
\hline
\multicolumn{5}{l}{\textbf{Center 1: New York University Langone Health (NYU)}}\\
\hline
Centralized&0.8665$\pm$0.0337 & 0.7709$\pm$0.1225&0.8350$\pm$0.0770 & 0.8754$\pm$0.0389\\
FedAvg~\cite{mcmahan2017communication}&0.8128$\pm$0.0445 & 0.6842$\pm$0.1305&0.8743$\pm$0.0212 & 0.9005$\pm$0.0479\\
FedProx ($\mu=0.1$)~\cite{li2020federated}&0.8195$\pm$0.0531 & 0.6892$\pm$0.1456&0.8540$\pm$0.0311 & 0.8898$\pm$0.0536\\
FedProx ($\mu=0.3$)~\cite{li2020federated}&0.7319$\pm$0.1591 & 0.7019$\pm$0.1482&0.8679$\pm$0.0314 & 0.9267$\pm$0.0360\\
\hline
\multicolumn{5}{l}{\textbf{Center 2: Mayo Clinic Florida (MCF)}}\\
\hline
Centralized&0.6571$\pm$0.0849 & 0.7480$\pm$0.1304&0.7244$\pm$0.1266 & 0.8146$\pm$0.1011\\
FedAvg~\cite{mcmahan2017communication}&0.6495$\pm$0.0986 & 0.7431$\pm$0.1005&0.7079$\pm$0.0494 & 0.8085$\pm$0.0825\\
FedProx ($\mu=0.1$)~\cite{li2020federated}&0.6575$\pm$0.1113 & 0.7366$\pm$0.1214&0.6780$\pm$0.0871 & 0.8075$\pm$0.0988\\
FedProx ($\mu=0.3$)~\cite{li2020federated}&0.6502$\pm$0.1348 & 0.7494$\pm$0.1322&0.6387$\pm$0.1016 & 0.8110$\pm$0.0657\\
\hline
\multicolumn{5}{l}{\textbf{Center 3: Northwestern University (NU)}}\\
\hline
Centralized&0.8496$\pm$0.0582 & 0.6766$\pm$0.1041&0.8293$\pm$0.0911 & 0.5734$\pm$0.1699\\
FedAvg~\cite{mcmahan2017communication}&0.9033$\pm$0.0236 & 0.7466$\pm$0.1361&0.9089$\pm$0.0243 & 0.5794$\pm$0.1795\\
FedProx ($\mu=0.1$)~\cite{li2020federated}&0.8815$\pm$0.0252 & 0.7346$\pm$0.1093&0.8493$\pm$0.1016 & 0.6018$\pm$0.2242\\
FedProx ($\mu=0.3$)~\cite{li2020federated}&0.8816$\pm$0.0246 & 0.7354$\pm$0.1427&0.8929$\pm$0.0220 & 0.5824$\pm$0.2704\\
\hline
\multicolumn{5}{l}{\textbf{Center 4: Allegheny Health Network (AHN)}}\\
\hline
Centralized&0.8750$\pm$0.1250 & 0.6667$\pm$0.4082&0.6750$\pm$0.1920 & 0.7083$\pm$0.2976\\
FedAvg~\cite{mcmahan2017communication}&0.5000$\pm$0.1768 & 0.3333$\pm$0.2357&0.7625$\pm$0.1781 & 0.8542$\pm$0.1488\\
FedProx ($\mu=0.1$)~\cite{li2020federated}&0.5625$\pm$0.3248 & 0.5833$\pm$0.4330&0.7000$\pm$0.2761 & 0.7083$\pm$0.1816\\
FedProx ($\mu=0.3$)~\cite{li2020federated}&0.5625$\pm$0.3698 & 0.5833$\pm$0.3632&0.8875$\pm$0.1139 & 0.7917$\pm$0.2165\\
\hline
\multicolumn{5}{l}{\textbf{Center 5: Mayo Clinic Arizona (MCA)}}\\
\hline
Centralized&0.6667$\pm$0.2041 & 0.7326$\pm$0.2388&0.5500$\pm$0.2021 & 0.7500$\pm$0.2500\\
FedAvg~\cite{mcmahan2017communication}&0.7500$\pm$0.1443 & 0.8507$\pm$0.1588&0.4833$\pm$0.0957 & 0.7500$\pm$0.1531\\
FedProx ($\mu=0.1$)~\cite{li2020federated}&0.7500$\pm$0.1443 & 0.8819$\pm$0.1185&0.4417$\pm$0.1140 & 0.6875$\pm$0.2724\\
FedProx ($\mu=0.3$)~\cite{li2020federated}&0.6250$\pm$0.1382 & 0.8819$\pm$0.1185&0.3917$\pm$0.0682 & 0.8125$\pm$0.1083\\
\hline
\multicolumn{5}{l}{\textbf{Center 6: Istanbul University Faculty of Medicine (IU)}}\\
\hline
Centralized&0.7812$\pm$0.0699 & 0.6450$\pm$0.2495&0.7948$\pm$0.0487 & 0.8336$\pm$0.1106\\
FedAvg~\cite{mcmahan2017communication}&0.7969$\pm$0.0518 & 0.6302$\pm$0.2020&0.7635$\pm$0.0900 & 0.7879$\pm$0.1325\\
FedProx ($\mu=0.1$)~\cite{li2020federated}&0.7500$\pm$0.0765 & 0.5946$\pm$0.2063&0.7958$\pm$0.0909 & 0.7742$\pm$0.1358\\
FedProx ($\mu=0.3$)~\cite{li2020federated}&0.7031$\pm$0.1945 & 0.6254$\pm$0.1640&0.7781$\pm$0.0285 & 0.7986$\pm$0.1147\\
\hline
\multicolumn{5}{l}{\textbf{Center 7: Erasmus Medical Center (EMC)}}\\
\hline
Centralized&0.8322$\pm$0.0776 & 0.8445$\pm$0.0730&0.7482$\pm$0.1306 & 0.8395$\pm$0.1171\\
FedAvg~\cite{mcmahan2017communication}&0.8079$\pm$0.0202 & 0.6687$\pm$0.1754&0.7720$\pm$0.1172 & 0.6875$\pm$0.1210\\
FedProx ($\mu=0.1$)~\cite{li2020federated}&0.7947$\pm$0.0376 & 0.8331$\pm$0.1508&0.7720$\pm$0.0684 & 0.8174$\pm$0.1610\\
FedProx ($\mu=0.3$)~\cite{li2020federated}&0.7816$\pm$0.0262 & 0.8112$\pm$0.1536&0.7714$\pm$0.0606 & 0.7868$\pm$0.1132\\
\hline
\multicolumn{5}{l}{\textbf{Global}}\\
\hline
Centralized&0.7990$\pm$0.0152 & 0.7813$\pm$0.0206&0.7823$\pm$0.0509 & 0.8237$\pm$0.0102\\
FedAvg~\cite{mcmahan2017communication}&0.7929$\pm$0.0204 & 0.7619$\pm$0.0102&0.8108$\pm$0.0169 & 0.8124$\pm$0.0298\\
FedProx ($\mu=0.1$)~\cite{li2020federated}&0.7852$\pm$0.0123 & 0.7646$\pm$0.0121&0.7829$\pm$0.0287 & 0.8135$\pm$0.0177\\
FedProx ($\mu=0.3$)~\cite{li2020federated}&0.7526$\pm$0.0490 & 0.7667$\pm$0.0109&0.7927$\pm$0.0241 & 0.8181$\pm$0.0215\\
\hline
\end{tabular}
\label{tab: fl}
\end{center}
\end{table*}

\subsection{Multi-Center IPMN MRI Binary Classification}
In this experiment, we apply 4-fold cross-validation due to the limited number of high-risk cases (only 4) available at Center 4 (AHN). The training process is the same as the one in the previous 3-class classification experiments.

Table~\ref{tab: fl} presents multi-center IPMN MRI binary classification results. Here, global ACC is computed as a weighted average of the ACC for each center, with weights proportional to the number of test images per center. Global AUC, on the other hand, is calculated by pooling all test labels and prediction scores across centers rather than by just averaging the AUC values from each center. This is because AUC is influenced by the overall distribution of true labels and predicted probabilities across the dataset.

\section{Conclusion}
In this study, we focused on classifying IPMN cases using multi-center pancreas MRI datasets and advanced deep learning techniques. By employing both a 3-class and a targeted 2-class classification approach, we aimed to accurately identify high-risk cases, which is crucial for effective clinical decision-making. Our findings highlight the performance of various 3D deep neural networks, supported by robust evaluation methods such as 4-fold cross-validation and the AdamW optimizer. The results emphasize the potential of federated learning to harness diverse datasets while preserving data privacy, laying the groundwork for future research aimed at enhancing classification accuracy and improving patient outcomes in the management of IPMN.

\section{Compliance with ethical standards}
\label{sec:ethics}
This study was performed in line with the principles of the Declaration of Helsinki. Approval was granted by Northwestern University (No. STU00214545).
% \begin{itemize}
%   \item ``This is a numerical simulation study for which no ethical
%     approval was required.'' 
%   \item ``This research study was conducted retrospectively using
%     human subject data made available in open access by (Source
%     information). Ethical approval was not required as confirmed by
%     the license attached with the open access data.''
%     \item ``This study was performed in line with the principles of
%       the Declaration of Helsinki. Approval was granted by the Ethics
%       Committee of University B (Date.../No. ...).''
% \end{itemize}

\section{Acknowledgments}
\label{sec:acknowledgments}
This work was supported by the following grants: NIH NCI R01-CA246704, R01-CA240639, U01-CA268808, NIH/NHLBI R01-HL171376, and NIH/NIDDK \#U01 DK127384-02S1.

\bibliographystyle{IEEEbib}
\bibliography{strings,refs}

\end{document}